\documentclass[aps,pra,superscriptaddress,twocolumn,a4paper,%
amsmath,amssymb,showpacs,floatfix]{revtex4}
\usepackage{bbm}
\usepackage{graphicx}
\usepackage{epstopdf}

\newcommand{\Fkt}[1]{\,\mathsf {#1}}

\newcommand{\e}{\Fkt{e}}
\newcommand{\eps}{\varepsilon}
\newcommand{\sumint}{\mathop{\hbox to -0.5ex{$\sum$}\int}}
\newcommand{\sumints}{\mathop{\hbox to -0.5ex{$\scriptstyle\sum$}\int}}

\begin{document}

\title{Making ultracold  molecules   in a two color pump-dump photoassociation  
scheme using chirped pulses} 

\date{\today}
\author{Christiane P. Koch}
\email{ckoch@fh.huji.ac.il}
\affiliation{Laboratoire Aim\'e Cotton, CNRS, B\^{a}t. 505, Campus d'Orsay,
91405 Orsay Cedex, France}
\affiliation{Department of Physical Chemistry and
The Fritz Haber Research Center, 
The Hebrew University, Jerusalem 91904, Israel}
\author{Eliane Luc-Koenig}
\affiliation{Laboratoire Aim\'e Cotton, CNRS, B\^{a}t. 505, Campus d'Orsay,
91405 Orsay Cedex, France}
\author{Fran\c{c}oise Masnou-Seeuws}
\affiliation{Laboratoire Aim\'e Cotton, CNRS, B\^{a}t. 505, Campus d'Orsay,
91405 Orsay Cedex, France}


\begin{abstract}
This theoretical paper investigates the formation of ground state molecules
from ultracold cesium atoms in a two-color scheme. Following previous work
on photoassociation with chirped picosecond pulses [Luc-Koenig et al.,
Phys. Rev. A {\bf 70}, 033414 (2004)], we investigate stabilization by a
second (dump) pulse.
By appropriately choosing the dump pulse parameters and time delay with respect to the
photoassociation pulse, we show that a large number of deeply bound molecules
are created in the ground triplet state. We discuss (i)
broad-bandwidth dump pulses which maximize the probability to form molecules
while creating a broad vibrational distribution as well as (ii) narrow-bandwidth
pulses populating a single vibrational ground state  level, bound by 113 cm$^{-1}$.
The use of chirped pulses makes the two-color scheme  robust, simple and efficient.
\end{abstract}

\maketitle
%--------------------------------------------------------------------------------%
\section{Introduction}
\label{sec:intro}
 
The field of cold molecules is rapidly developing, since  many research groups   
are working at the production of dense samples of cold ($T \le $1 K) or ultracold 
($T \le$ 100 $\mu$K) molecules \cite{doyle2004}. Moreover, the field of molecular condensates has 
opened recently, with experimental evidence for  molecules formed in a degenerate gas, 
starting either from an atomic Bose-Einstein 
condensate \cite{donley2002,claussen2002,herbig2003,xu2003,durr2003}, 
or from an atomic Fermi gas
\cite{jochim2003a,greiner2003,zwierlein2003,strecker2003,cubizolles2003,%
jochim2003b,regal2003}.
This intense activity is motivated by possible 
applications to very precise measurements (lifetimes, checks on parity violation, 
determination of a possible dipole moment of the electron as a check for the standard 
model in elementary particles), Bose-Einstein condensation of complex systems (molecule laser), 
or the emergence of a new ultracold chemistry. 
   
Among the various routes to produce ultracold molecules, photoassociation has 
emerged as a purely optical technique starting from an assembly of laser-cooled 
atoms in a trap.
The reaction takes place when a pair of ground state atoms interacting via the  
potential $V_\mathrm{ground}$ absorbs a photon red-detuned compared to the atomic line.   
At resonance,  one vibrational level in  an  excited electronic potential $V_\mathrm{exc}$ 
of the molecule is populated. The reaction can be interpreted as a vertical transition 
at large distances, forming a long range molecule. We shall consider the example of 
two cold cesium atoms colliding in the lower triplet  $a^3 \Sigma_u^+(6s+6s)$ potential
and forming a  molecule   in a vibrational level $v'$ of the  $0_g^-(6s+6p_{3/2})$ 
potential.  The first experimental observation of ultracold Cs$_2$ molecules has
made use of this reaction \cite{fioretti98}.
 
In order to make  molecules in the \textsl{ground} or \textsl{lower triplet 
state}, the photoassociation step must be followed  by a stabilization step. The  
photoassociated molecules are short-lived and   decay via spontaneous emission,  
most often through  a vertical transition at large interatomic distances $R$
giving back    a pair of atoms. However, specific mechanisms which 
favor radiative decay at shorter distances exist and allow for populating bound  vibrational  
levels in the potential  $V_\mathrm{ground}(R)$ (or possibly, for instance in case of 
heteronuclear dimers, to another potential correlated to the same
asymptote~\cite{SagePRL05}). 
In the special case  of  the double well $0_g^-(6s+6p_{3/2})$ excited potential
of Cs$_2$, the smooth barrier between the two wells acts as a ``speed bump'' and slows 
down the vibrational motion in the external well such that vertical transitions 
can take place at intermediate distances $R_\mathrm{int}$. 
Stable molecules formed by spontaneous 
emission through this mechanism have been detected in a MOT \cite{fioretti98}. Other 
efficient stabilization mechanisms such as resonant coupling \cite{dion01,dulieu03,kerman04}
exist.

In a condensate, in order to retain coherence, spontaneous emission should be avoided 
and a stabilization step involving stimulated emission should be implemented  within a 
two-color experiment. However, with continuous wave (cw) lasers such experiments are hardly 
feasible since the scheme is fully reversible:  Population is pumped back from bound levels 
of the ground state to the excited state, and then to   continuum levels of the ground state, 
i.e. the molecule is dissociated. 
The use of laser pulses  becomes thus unavoidable.

In previous work \cite{vala01,luc2004a,luc2004b}, the possibility to increase the 
efficiency of the photoassociation reaction by use of chirped laser pulses was considered. 
The frequency of such pulses has a linear time-dependence. The parameters of the pulse 
were optimized to achieve a total transfer of population under adiabatic following 
conditions \cite{cao98,cao00,WrightPRL05}. %In the example described above, 
It was shown \cite{luc2004a,luc2004b} that  
a significant improvement of the photoassociation rate  compared to 
cw excitation can be obtained, 
since during the pulse the resonance condition is fulfilled for several 
vibrational levels $v'$.
However, the quantity which must be 
improved is the formation rate of ground state molecules 
rather than the photoassociation rate.  One advantage of excitation 
with chirped pulses is that the chirp parameter can be chosen to modify the shape of the 
vibrational wave packet formed in the  potential $V_\mathrm{exc}$.  This may improve the 
efficiency of the stabilization step, and thus the formation rate of ground state molecules.

The aim of the present work is therefore to suggest a two-color experiment using 
short laser 
pulses and to investigate how efficiently a second pulse can transfer the population 
to the bound vibrational levels of the ground state. Starting from the  vibrational 
wave packet formed in the excited state by photoassociation, 
we introduce a second pulse which is delayed in time from the first pulse. The time delay
is chosen   such  that the wave packet has reached the 
vicinity of the inner turning point, where a vertical radiative 
transition to the ground state  is expected to be efficient. The parameters of the 
second pulse, as well as the time delay between the two pulses
are varied to improve the efficiency of the 
transfer, with two different goals: to either maximize the total population transfer, 
or to selectively populate  one vibrational level. The paper is organized as follows:
In Sec. \ref{sec:previous} we recall the main results of previous work on 
photoassociation with chirped laser pulses. The two-color scheme is introduced 
in Sec. \ref{sec:2color}, and the best choice for the dump pulse discussed by considering   
the Franck-Condon overlap between  the vibrational wave packet in the excited state and 
wave functions of the bound levels in the ground state. The results for maximum population
 transfer are
presented in Sec. \ref{sec:many} while Sec. \ref{sec:single} reports on the
selective population  of a single level. Finally Sec. \ref{sec:concl} concludes.
 
%--------------------------------------------------------------------------------%
\section{Chirped Pulse Photoassociation in  {C{s}}$_2$  and creation of a focussed 
vibrational wave packet}
\label{sec:previous}
We  briefly summarize the main results of Refs. \cite{luc2004a,luc2004b} for the 
photoassociation  of cold cesium atoms by chirped laser pulses. The main difference 
to cw excitation lies in two facts: Several levels can be  populated in the 
photoassociation step, and  the vibrational wave packet can be shaped in order to 
allow for spatial "focussing", as described below.
We start from a  transform-limited pulse with Gaussian envelope, hereafter referred 
to as pump pulse,
with maximum at $t=t_P$, and with full width at half maximum of  the temporal 
intensity profile (FWHMI) $\tau_L$=15 ps. Chirping the pulse stretches the 
FWHMI to $\tau_C$=34.8 ps and decreases the maximum intensity 
$I_L  \to I_L \frac{\tau_L}{\tau_C}$ ($I_L$ =120 kW cm$^{-2}$), conserving 
the energy carried by the field. Moreover,  98\% of this energy is carried during  
the  time window $[t_P-\tau_C,t_P+\tau_C]$. We consider the reaction
%\begin{equation}
\begin{eqnarray}
&&2\mathrm{Cs}(6s,F=4) + \hbar[\omega_{L}+\chi_P (t-t_P)]
 \rightarrow \\ \nonumber 
&& \quad \quad \quad 
\mathrm{Cs}_{2}(0_g^-(6s\,\,^{2}S_{1/2}+6p\,\,^{2}P_{3/2};v^{\prime},J'=0))
\end{eqnarray}
% \end{equation}
where the carrier frequency $\omega_{L}$ is chosen red-detuned relative to the 
D$_2$ atomic line, at resonance with the $v'$=$v'_L$=98 level in the external well of 
$V_{\mathrm{exc}}$ which is bound by 2.65 cm$^{-1}$. $\chi_P$ is the linear chirp 
rate in time domain, chosen negative and equal to -0.025 cm$^{-1}$ ps$^{-1}$. 
The parameters of the pump pulse, labelled ${\cal{P}}_-$ are recalled in Table~\ref{tab:pump}.
We will also consider  other possibilities for the pump pulse, 
with positive or zero chirp, or with a smaller detuning: such pulses are 
respectively labelled ${\cal{P}}_+$, ${\cal{P}}_0$ and ${\cal{P}}_-^{122}$. 
The latter pulse has a central frequency  resonant with the excited level $v'$=122. 
The parameters for ${\cal{P}}_-^{122}$ have been chosen to ensure the focussing condition 
without populating continuum levels in the photoassociation step.
\begin{table*}[htb]
\vspace{0.5cm}
\begin{tabular}{|c|c|c|c|c|c|c|c|c|c|}
\hline\noalign{\smallskip}
label& $\delta_L^{at}$ & $I_L$ & $\hbar \delta \omega$& $\tau_L$ & $\tau_C$ &$\hbar \chi_P$ &$\tau^P$&$f_P$&10$^4$$P_{\mathrm{exc}}$\\
\noalign{\smallskip}\hline\noalign{\smallskip}
${\cal{P}}_-$ &2.656 cm$^{-1}$ & 120 kW cm$^{-2}$&0.98 cm$^{-1}$&15 ps&34.8 ps&-0.025 cm$^{-1}$ps$^{-1}$&21.2 ps&2.32&3.2 \\
\noalign{\smallskip}\hline\noalign{\smallskip}
${\cal{P}}_+$ &2.656 cm$^{-1}$ & 120 kW cm$^{-2}$&0.98 cm$^{-1}$&15 ps&34.8 ps&+0.025 cm$^{-1}$ps$^{-1}$&21.2ps&2.32&3.5 \\
\noalign{\smallskip}\hline\noalign{\smallskip}
${\cal{P}}_0$ &2.656 cm$^{-1}$ & 120 kW cm$^{-2}$&0.98 cm$^{-1}$&15 ps&15 ps&0&21.2 ps&1&0.4 \\
\noalign{\smallskip}\hline\noalign{\smallskip}
${\cal{P}}^{122}_-$ &0.675cm$^{-1}$ & 120 kW cm$^{-2}$&0.26 cm$^{-1} $&57.5 ps&110 ps&-0.002 cm$^{-1}$ ps$^{-1}$&81.3 ps&1.91  &20\\
\noalign{\smallskip}\hline
\end{tabular}
\vspace{0.5cm} 
\caption{%
  Parameters for the main  pump pulse ${\cal{P}}_-$ considered  in
  Refs. \cite{luc2004a,luc2004b}: 
  detuning $\delta_L^{at}$, maximum intensity $I_L$, energy range associated to the spectral 
  width $\hbar \delta \omega$, temporal widths (FWHMI) $\tau_L$ and $\tau_C$, linear chirp 
  parameter $\chi_P$; new parameters used in the present work are the temporal
  width (FWHME) before 
  chirping $\tau^P$, and stretching factor $f_P$ (see Eq.~(\ref{eq:newparam}) in text). Also 
  indicated are other pump pulses with different chirp or different detuning. In the last column, 
  we indicate the population transferred to the excited state by the photoassociation pulse.
}
\label{tab:pump}
\end{table*}
Note that in order to compare to earlier papers, we have also considered another the 
temporal  width,
\begin{eqnarray}
\tau^P =  \sqrt{2} \tau_L,\nonumber \\
\tau^P_C=\sqrt{2} \tau_C, \nonumber \\
 f_P =\tau_C/\tau_L=\tau^P_C/\tau^P,
\label{eq:newparam}
\end{eqnarray}
where $\tau^P $ is now defined as the full width at half maximum of the electric field 
profile (FWHME) of the transform-limited pulse, while in Refs.~\cite{luc2004a, luc2004b} 
$\tau_L$  was defined with respect to intensity (FWHMI). $f_P$ is the stretching factor
due to chirping. During the time window, the central  frequency is swept  
from $\omega_L - \chi \tau_C$ to $\omega _L+ \chi \tau_C$ such that the  resonance 
condition is satisfied  for 15 vibrational levels around $v'_L$=98 in the energy range 
$2\hbar \mid \chi \mid \tau_C=2\times 0.87$~cm$^{-1}$. Depending upon the sign of 
the chirp, the frequency is increasing ($\chi_P >0$) or decreasing ($\chi_P <0$) with time. 

The dynamics of the photoassociation process was studied by numerical solution of 
the time-dependent Schr\"odinger equation describing the motion of  wave packets 
in the ground state potential  $a^3 \Sigma_u^+(6s+6s)$ ($V_{\mathrm{ground}}$) and in  the 
$0_g^-(6s\,\,^{2}S_{1/2}+6p\,\,^{2}P_{3/2})$ excited potential ($V_{\mathrm{exc}}$)
coupled by the electromagnetic field. The initial state is represented by a 
stationary continuum wave function describing the relative motion ($s$-wave) of  two colliding 
Cs (6s, F=4) atoms, with an initial kinetic energy of 54~$\mu$K. The wave packet in the 
excited potential 
$V_{\mathrm{exc}}$ is analyzed by projection onto the vibrational wave functions of 
the $v'$ levels in the external well.
The main results of the calculations are:
\begin{itemize}
\item During the pulse, many vibrational levels are populated. However, after the pulse, the 
  population remains only in the  15 vibrational levels resonantly excited around $v'_L$=98. 
  This defines a photoassociation window in the energy domain, for the levels located between 
  $v^{\prime}_1$=106 and $v^{\prime}_2$ = 92, bound by  1.74 cm$^{-1}$ and  
  3.57 cm$^{-1}$, respectively.
\item In the impulsive approximation \cite{banin94}, assuming that the relative motion 
  of the two atoms is frozen during the pulse, it is possible to analyze the numerical 
  results by a two-state model at each internuclear distance $R$. While the frequency of 
  the laser is swept from $\omega_L - \chi_P \tau_C$ to $\omega _L+ \chi_P \tau_C$, the crossing 
  point $R_c (t)$ of the two field-dressed potential curves is moving in the  
  $[R_\mathrm{min},R_\mathrm{max}]$ range. Assuming that the resonance condition 
  corresponds to a vertical transition at $R=R_c (t)$, this  defines a photoassociation 
  window in the $R$ domain:   in the example chosen,  $R_\mathrm{min}$ = 85 and 
  $R_\mathrm{max}$ =107 a$_0$. Within this range of distances, all the population of the 
  initial continuum level can be adiabatically transferred to the excited state. 
\item In the excited state, the levels $v^{\prime}$ which are populated have a classical 
  outer turning point $R^{v'}_\mathrm{out}$ in  the  $[R_\mathrm{min},R_\mathrm{max}]$ range, and 
  the two-state model predicts that the population transfer occurs around the time $t_{v'}$ 
  when $R^{v'}_\mathrm{out}$= $R_c (t_{v'})$. For a negative chirp, the position of the crossing 
  point varies from $R_\mathrm{max}$ to $R_\mathrm{min}$, so that the upper levels are 
  populated first (see Fig. \ref{fig:vib}). For the pump pulse considered, the level 
  $v^{\prime}_1$= 106 is populated at 
  the beginning of the time window ($t_1\sim (t_P-\tau_C$), the level $v^{\prime}_2$ = 92 
  is populated 
  at the end ($t_2 \sim t_P+\tau_C$), i.e. 70 ps later. It is the opposite for a positive chirp, 
  choosing the ${\cal P}_+$ pulse for photoassociation.
\item The total probability of population transfer to the excited state is reported in the 
  last column of Table~\ref{tab:pump}. The absolute value is discussed in 
  Refs.~\cite{luc2004a,luc2004b}, the important result for the present paper being that the 
  population obtained with  ${\cal{P}}_0$ is increased by a factor 8 when a chirped pulse 
  ${\cal{P}}_-$ is considered, and  by a factor of 6 when the pulse ${\cal{P}}_-^{122}$, 
  with a smaller detuning,  is used.   
\item After the end of the pulse, the vibrational wave packet $\Psi_{\mathrm{exc}}(R,t)$ 
  in the excited state is  moving from the outer classical turning point toward shorter 
  distances (see Fig.\ref{fig:vib}). We analyze it through  partial wave packets, with main 
  components corresponding 
  to  the various levels $v'$.  Each one is reaching the inner turning point  at a time 
  which depends upon the excitation time $t_{v'}$ and the vibrational period $T_{vib}(v')$. 
  In the photoassociation window considered here, the vibrational periods are in the range 
  350-196 ps, yielding $\frac{1}{2}T_\mathrm{vib}(v^{\prime}_1)$ = 175 ps and
  $\frac{1}{2} T_\mathrm{vib}(v^{\prime}_2)$ = 98 ps, respectively.  
  For a negative chirp,  we may  choose 
  the parameters of the pulse such as to create partial vibrational wave packets 
  which reach the inner 
  turning point approximately at the same time  
  $t_3 \approx t_1 +1/2 T_\mathrm{vib}(v^{\prime}_1)
  \approx t_2 +1/2 T_\mathrm{vib}(v^{\prime}_2)$.
  \begin{figure}[tb]
    \includegraphics[width=0.9\linewidth]{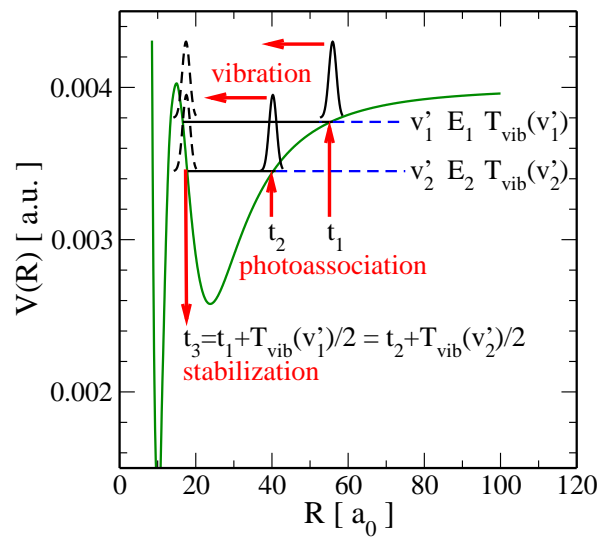}\hfill%
    \caption{(Color online) Photoassociation with a chirped pulse:
      For negative chirp, the level $v'_1$ is populated at the beginning ($t_1$), 
      while $v'_2$ is populated at the end ($t_2$) of the time window characterizing the 
      photoassociation pulse. The magnitude of the chirp
      can be chosen such that, after half a vibration period $T_{vib}$,
      the partial wave packets arrive at the potential barrier at the same time $t_3$.
      This time 
      ($t_3$) is optimal  for radiative stabilization, with a second pulse,  toward bound 
      ground state levels.
    }
    \label{fig:vib}
  \end{figure} 
  For long range molecules, the vibrational 
  period is known to obey  a scaling law \cite{stwalley73}, and it is easy to choose 
  the chirp parameter such that all the partial  wave packets in the 
  intermediate levels $ v^{\prime}_2 \le v' \le v'_1$ satisfy \cite{luc2004b}
  \begin{equation}
    \label{eq:focus}
    t_{v'}(\chi_P) +\frac{T_\mathrm{vib}(v^{\prime})}{2}\approx t_P +
    \frac{T_\mathrm{vib}(v'_L)}{2}=t_3.
  \end{equation}
  Focussing of the wave packet is then obtained at time $t_3$, after half a vibrational period. 
  In contrast, for a positive chirp the lower levels are populated first, $t_2 < t_1$, 
  and a spreading of the vibrational wave packet is expected. 
\item
  Several authors in the past have suggested ways of designing laser fields for the
  generation of spatially squeezed molecular wave packets \cite{averbukh93,abrashkevich94}. 
  Since they considered low vibrational levels, a numerical optimization procedure 
  appeared to be necessary. Here, we are dealing with long range molecules, where the motion is 
  controlled by the asymptotic potential, and the availability of scaling laws makes the 
  optimization easier, as for atomic Rydberg wave packets \cite{ahn2000}.
\item
  We should note that in the chosen example, the vibrational progression and the scaling law 
  of the vibrational period are slightly perturbed due to a tunneling effect for the levels 
  in the vicinity of  $v^{\prime}$=96 \cite{vatasescu02,luc2004a}. This  leads to a focussing 
  time $t_3$=$t_P$+135 ps, 
  slightly different from $\frac{1}{2}T_\mathrm{vib}(v'=98)$ =125 ps. 
  Moreover, the early population of levels which are perturbed by tunneling 
  leads to population transfer 
  to the inner well of the $0_g^-(6s+6p_{3/2})$ excited potential. 
  This tunneling effect is not a general phenomenon, but very  specific to the cesium $0_g^-$ 
  molecular curve, and to the value chosen for the detuning.
  A detailed investigation would require consideration of two coupled channels 
  in the region of the inner potential \cite{vatasescu02}, beyond the scope of this work. 
\end{itemize}

In the present calculations, we start from the wave packet in the excited state 
and investigate   how  the population can efficiently be transferred to vibrational 
levels of the ground state by the second pulse. As in 
our previous work, this is 
achieved by studying the wave packet dynamics by numerical solution of the time-dependent 
Schr\"odinger equation (cf. Ref.~\cite{luc2004a,luc2004b} for an overview over
numerical methods and Appendix~\ref{sec:box} for a discussion of the representation of
continuum states).

\section{The Two-color scheme}
\label{sec:2color}

\begin{figure*}[tb]
  \includegraphics[width=0.45\linewidth]{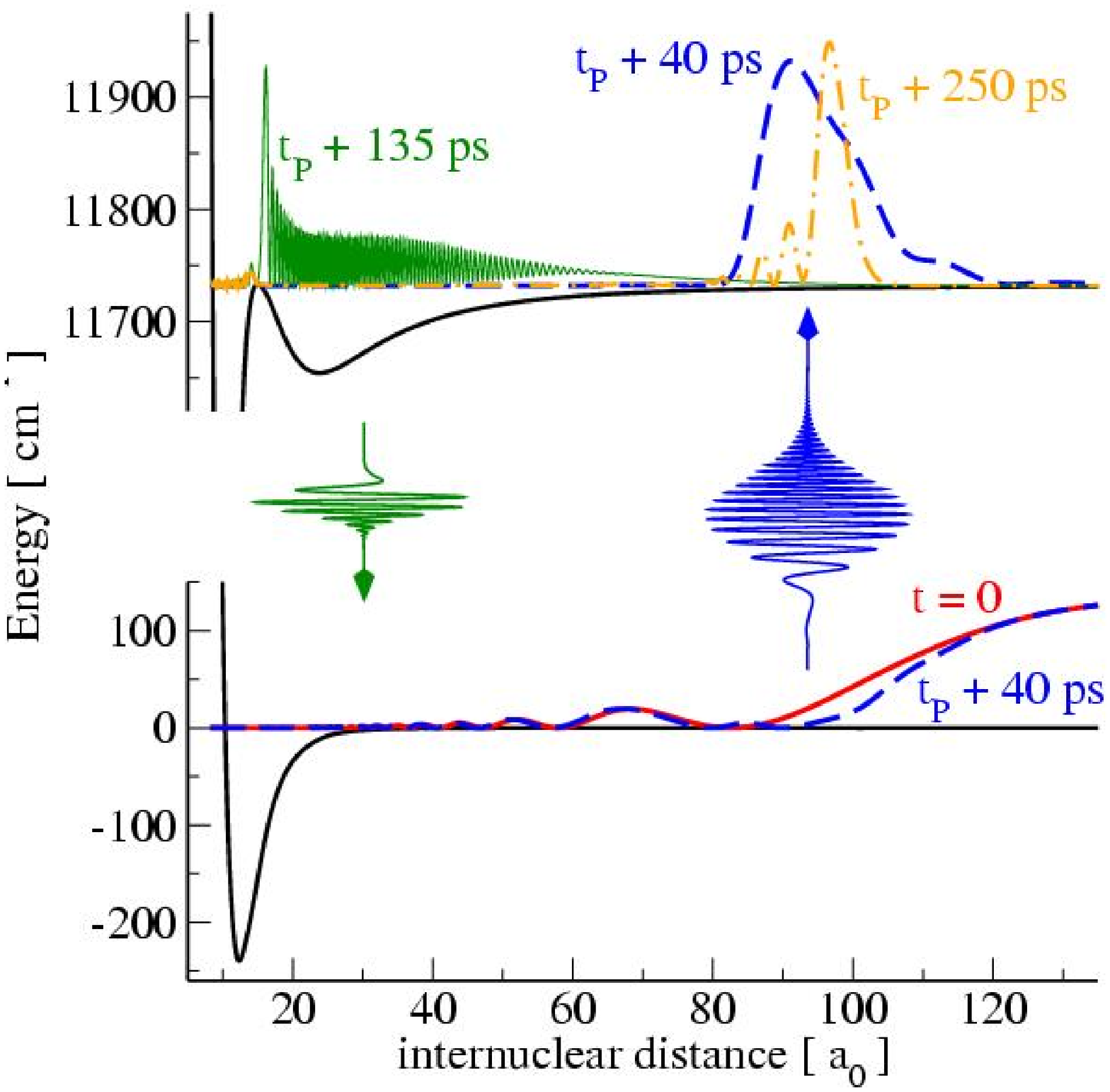}\hfill%
  \includegraphics[width=0.45\linewidth]{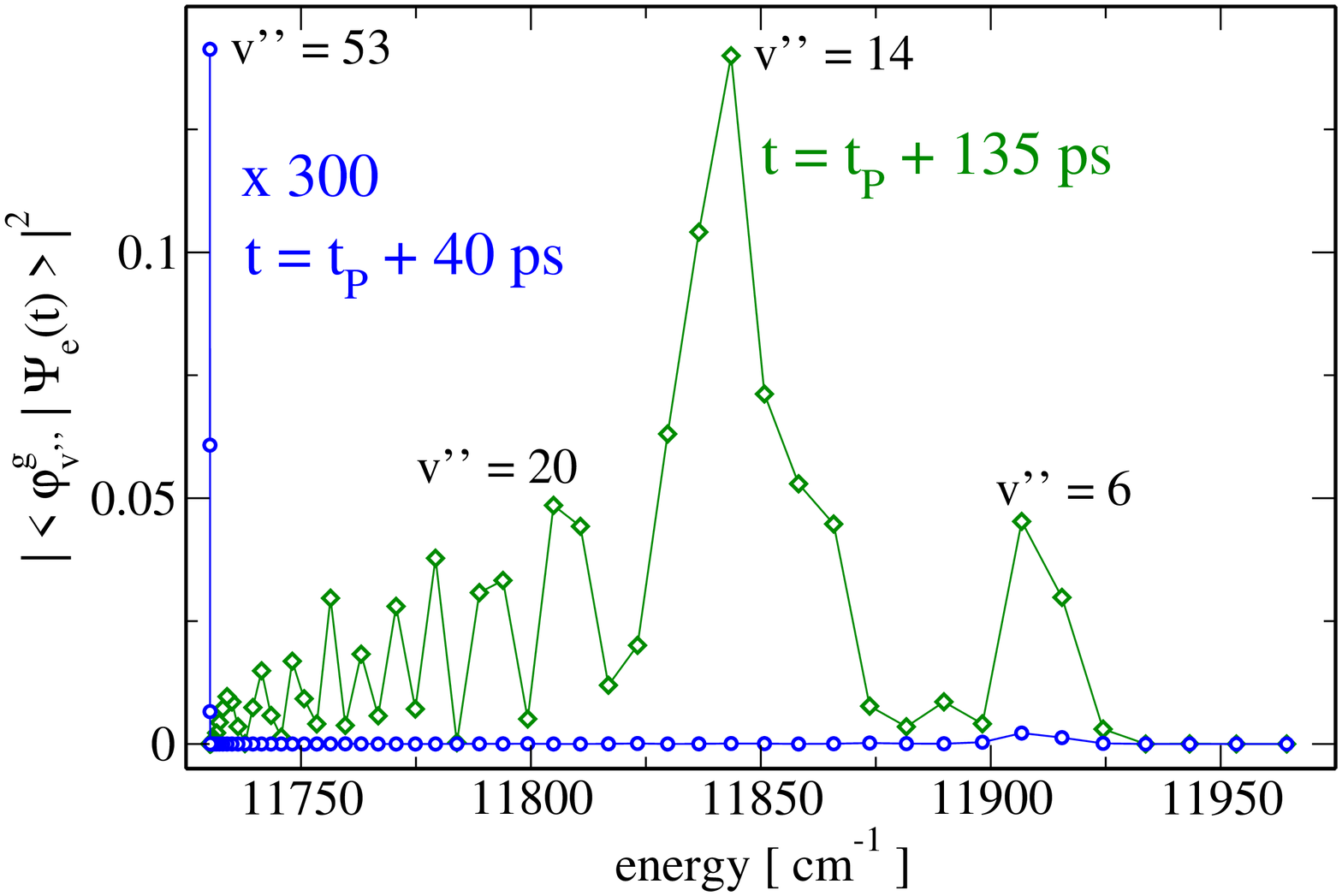}
  \caption {(Color online) left: The two-color scheme: A chirped picosecond pulse ${\cal{P}_-}$
    (see Table~\ref{tab:pump}) excites a wave 
    packet in  the electronically excited state. A second, shorter  pulse ${\cal{D}}$ 
    (see Tables~\ref{tab:dump_short}, \ref{tab:dump_long}) dumps this wave 
    packet to the electronic ground state.
    right: Franck-Condon overlap between the time-dependent wave packet 
    (population normalized to one, see text) in the excited state, and ground state 
    vibrational levels   at times $t=t_P $+40 ps, just after the photoassociation
    pulse (blue circles) 
    and at $t=t_P $+135~ps $\approx t_\mathrm{in}$ (green diamonds), when the wave packet is 
    focussed at the inner barrier; $t_P$  is the time when the photoassociation pulse is maximum. 
    When the wave packet is focussed at the inner turning point, there is a maximum 
    overlap with the level $v''=14$  of the ground state.
  }
  \label{fig:scheme}
\end{figure*}

The scheme we suggest 
is illustrated in Fig.~\ref{fig:scheme} (left), where we have 
represented, at different times, the wave packet created by the negatively 
chirped picosecond pulse ${\cal P}_-$. 
We consider a vibrational wave packet $\Psi_{\mathrm{exc}}(R,t=t_P + t_\mathrm{dyn})$,  
which  has evolved on the excited state potential for time $t_\mathrm{dyn}$ after 
the maximum of the pump pulse $t_P$.  
In order to discuss directly transfer probabilities, we will
renormalize the wave packet before applying the dump pulse, 
such that at any time 
$<\Psi_{\mathrm{exc}} |\Psi_{\mathrm{exc}}>$=1. The variation of the photoassociation 
probability as reported in the last column of Table~\ref{tab:pump} 
will be reconsidered in the conclusion. 

\subsection{The Franck-Condon overlap and its time-dependence: choice of the target 
vibrational level in the stabilization step} 
 We first compute  the Franck-Condon overlap matrix elements 
$|\langle\Psi_{\mathrm{exc}}(t_\mathrm{dyn})|\varphi^g_{v''} \rangle|^2$ of the 
excited state wave packet with the bound ground state eigenfunctions
(cf. Fig.~\ref{fig:scheme}, right). The results markedly depend upon $t_\mathrm{dyn}$,  
and so does the probability to form stable molecules. For instance, just after the 
photoassociation pulse, at time $t_P$+ 40 ps $\sim t_P+\tau_C$
(circles in Fig.~\ref{fig:scheme}, right), the Franck-Condon overlap factors 
are negligible for most bound levels of the ground state, except for  the last
level $v''=53$ which is very loosely bound (5 $\times$ 10$^{-6}$ cm$^{-1}$~\cite{luc2004a}). 
In contrast, half a vibrational period later, at time $t_P+$135~ps when the wave packet has 
moved to the inner turning point (diamonds in Fig.~\ref{fig:scheme}, right),
a significant overlap is obtained  with various deeply bound levels  of the ground state, 
in particular with $v''=6,$ 14, 20. 
\begin{figure}[tb]
  \includegraphics[width=0.85\linewidth]{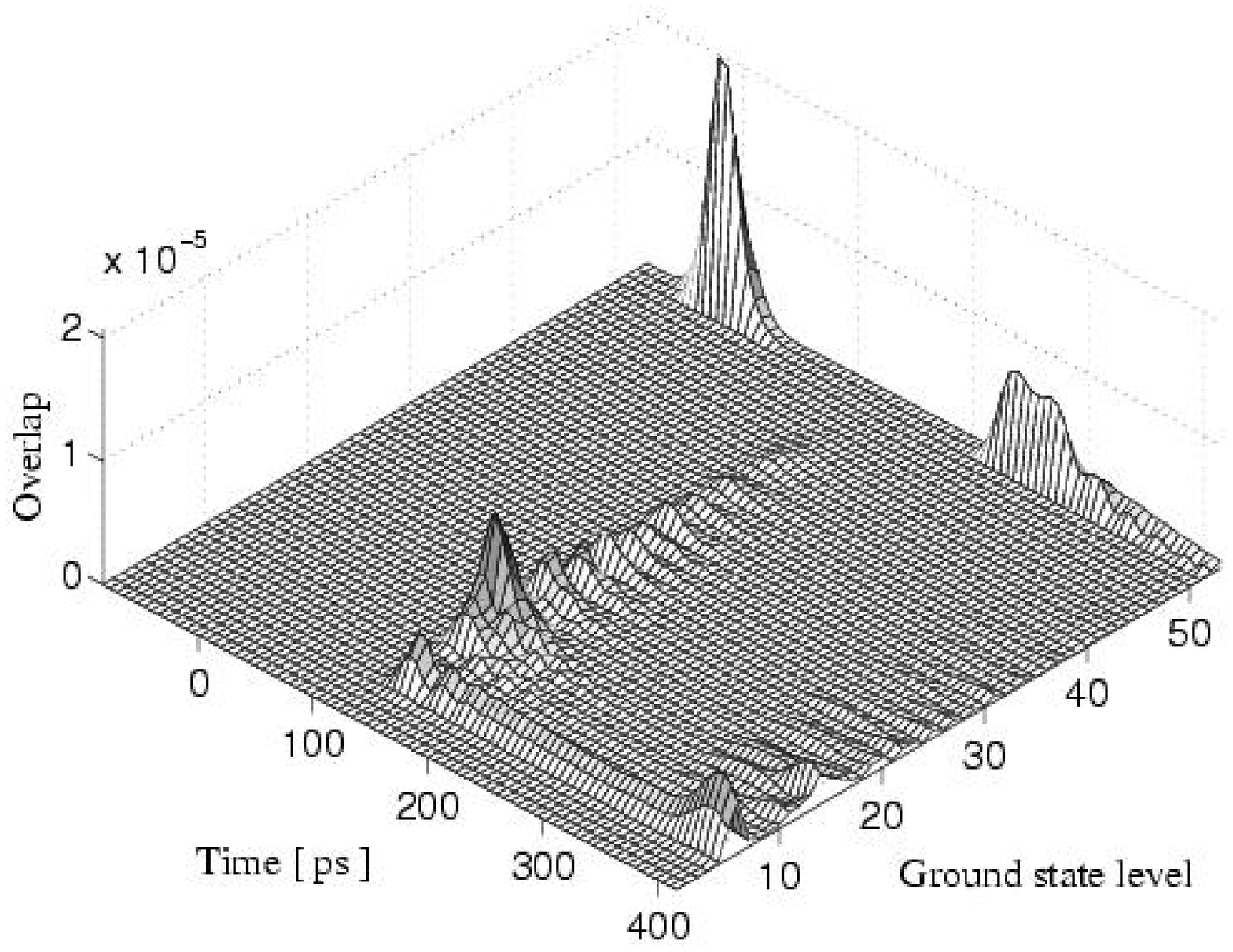}
  \caption {A two-dimensional plot of the Franck-Condon overlap of Fig.~\ref{fig:scheme}(b) 
    between the time-dependent wave packet in the excited state and all bound vibrational 
    levels of the ground state. $t_P$ corresponds to $t=0$. It is clear 
    that the overlap with deeply bound ground state levels (around $v''$=14) is maximum when 
    focussing of the excited state wave packet occurs,  135~ps or half a 
    vibrational period after $t_P$. In contrast, when the wave packet is at its
    outer turning point 
    just after the PA pulse and again at about $t_P+$250~ps (one vibrational period $T_{vib}$), 
    the  overlap with only the last bound levels is significant. 
    Moreover, tunneling to and trapping of population in the inner potential well lead
    to overlap with ground state levels around $v''=6$:  this is 
    specific to the cesium $0_g^-(P_{3/2})$ state and to the value chosen for the detuning. 
    Note the stability of this population during half a vibrational period.
  }
  \label{fig:schemenew}
\end{figure}
The large overlap with $\varphi^g_{v''=14}(R)$ can be explained by the coincidence 
of the inner classical turning point of the level $v^{\prime}$=92 and the outer 
turning point of $v''$=14, such that %the transition 
$v^{\prime}$=92 $\to$  $v''$=14 is a 
Franck-Condon transition. This large value is restricted to a short ($\sim\pm 2$~ps) 
time window  around $t_P+135$~ps as illustrated  in Fig.~\ref{fig:schemenew}, where 
the time-dependence of the Franck-Condon overlap %with the bound vibrational levels $v''$ 
is shown. 
Note that the overlap with $v''=6$ is due to population in the inner well of $V_\mathrm{exc}$
due to tunneling~\cite{vatasescu02}. 
From Fig.~\ref{fig:schemenew}, we can indeed see that this population remains trapped  
at the  considered timescale. 

The above mentioned large  overlap with   the last vibrational level of the 
ground state just after  the photoassociation pulse ($t-t_P \le \tau_C$)  is seen to  
reappear after one vibrational period, when the wave packet reaches again the outer 
turning point. However, we will not choose one of  the last bound levels of the 
ground state as our target vibrational level $v''$, since a significant  population 
is already achieved  in a one-color experiment~\cite{luc2004a,luc2004b}, making the 
implementation of a two-color  scheme superfluous. 
In fact, the specific advantage of a pump-dump experiment is the possibility of 
stabilization into  deeply bound levels in the ground state. We will therefore 
focus on populating the level $v''=14$ with binding energy $E_\mathrm{b} = 113$~cm$^{-1}$,
and  we will  consider a second  (or dump) pulse ${\cal D}$ with parameters  adapted to efficient 
population transfer toward  $v''=14$, in particular with a time delay close to 135 ps.
   
\subsection{Choice of  parameters for the second pulse}
\label{ssec:param}
For the dump pulse ${\cal{D}}$, we define a central frequency $\omega_L^D$, a 
maximum intensity $I_L$, a duration (FWHME) of the transform-limited pulse $\tau_D$, 
the time of  maximum amplitude $t_D$, and 
a stretching factor $f_D$ characterizing the strength of the chirp. 
The choice of the second color is dictated by the choice of the target levels.
For  $v'' = 14$ (cf. Fig.~\ref{fig:scheme}, right), we have chosen  a  central 
frequency $\omega_L^D=11843.5$~cm$^{-1}$,
 which is resonant with  $v''= 14$ for a binding energy of the excited state
wave packet equal to  2.1 cm$^{-1}$.  This binding energy, computed  as 
$\langle\Psi_{\mathrm{exc}}(R,t)|\hat{T}+V_{\mathrm{exc}}|\Psi_{\mathrm{exc}}(R,t)\rangle$, 
where $\hat{T}$ is the kinetic energy operator, stays time-independent once 
the photoassociation pulse is over. It is found 0.5 cm$^{-1}$ smaller than the 
binding energy  of the level $v'=v'_L=98$ which is  resonant with the central
frequency of the pump pulse. This  difference reflects   the non-symmetric 
population of the $v'$ levels relative to $v'_L$, the photoassociation process 
being much more efficient for the upper levels.

The spectral width of the second pulse $\cal{D}$ is limited by the requirement that mostly  bound 
levels should be populated in the ground state. Since  $v''$=14 is bound by 
113 cm$^{-1}$, the energy spread of the vibrational wave packet in the excited state 
can be neglected and the maximum spectral width  is
estimated as 2$\times$113~cm$^{-1}$. Therefore, for 
short pulses ${\cal{D}}$ (FWHME $\tau^D \le 93$~fs),
population transfer to the continuum is to be expected
leading to the formation of pairs of ``hot'' atoms.  
This  conclusion relies on overlap integrals and needs to be corrected by the possible energy 
dependence  of the transition dipole moment $\mu(R)$.
Finally, if we wish to selectively excite the level $v''=14$, the spectral width is limited 
by the vibrational energy splitting around $v''=14$ which is about 7~cm$^{-1}$. 
The typical pulse temporal width $\tau^D$ of the dump 
pulse should therefore be larger than 2 ps. 
In the following, we will discuss two schemes to form ground state molecules:
\begin{enumerate}
\item For short pulses in the 10-300 fs range, important  population transfer to 
  several bound levels of 
  the ground electronic state can be obtained (cf. Sec.\ref{sec:many}). The goal  is then to
  maximize the total population transfer. The intensity,  duration and possibly stretching 
  factor, i.e. chirp,  of the ${\cal {D}}$ pulses will be systematically varied.
\item For longer pulses in the picosecond domain, transfer of population to a 
  single  vibrational level, namely $v''$=14, is possible (see Sec.\ref{sec:single}). The goal
  is then to optimize selective population transfer to a chosen level.
\end{enumerate}

The third parameter is the time-delay between the two pulses, defined as $t_D-t_P$. 
The discussion can be guided by the following arguments, illustrated in 
Fig.~\ref{fig:probability}:
\begin{figure}[tb]  
  \includegraphics[width=0.95\linewidth]{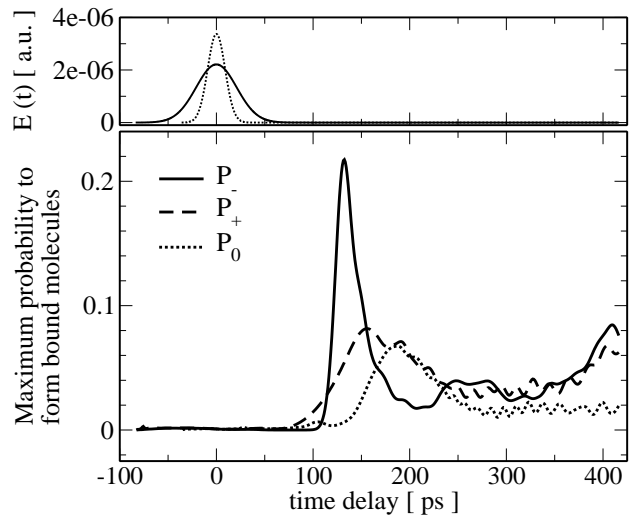}
  \caption{(top) The photoassociation pulses ${\cal P_{\pm}} $,${\cal P}_0 $. 
    (bottom) The upper limit to the probability of forming bound molecules
    is given by the sum over Franck-Condon
    overlap matrix elements between the time-dependent excited state wave packet 
    (assumed normalized to one)
    and all the bound levels of the ground state. The highest probability of about 20\% is 
    expected for the focussed wave packet (at time $t=t_P+135$~ps).  
    }
  \label{fig:probability}
\end{figure}  
\begin{itemize}
\item First, we can use a simple model for the choice of  the time delay $t_D -t_P$.  
  Considering  the wave packet $\Psi_{\mathrm{exc}}(R,t_D)$, it is possible to estimate 
  an upper limit to the probability of forming molecules from the sum over the Franck-Condon 
  overlap matrix elements $|\langle\Psi_{\mathrm{exc}}(R,t_D)|\varphi^g_{v''} \rangle|^2$. 
  After a short and weak dump pulse at time $t_D$, the wave packet created in the ground state is 
  estimated by 
  \begin{equation}
    \label{eq:psig_perturb}
    |\psi_g(t) \rangle = \sumint_E \mu \eps(E)
    \langle  \varphi_E^g|\psi_e(t_D)\rangle
    \e^{-\frac{i}{\hbar}E(t-t_D)}|\varphi_E^g\rangle ,
  \end{equation}
  where $\sumints_E$ denotes jointly the sum over bound levels with eigenenergies $E=E_i$
  and the integral over continuum states with energy $E$, while  $\eps(E=\hbar\omega)$
  is the Fourier transform of the field $\eps(t)$ and $\mu$ the dipole moment, assumed to 
  be $R-$independent.
  Eq.~(\ref{eq:psig_perturb}) implies that vibrational levels and continuum states
  which are resonant within the bandwidth $\eps(\omega)$ of the pulse are excited.
  The population transfer is proportional to the time-dependent Franck-Condon 
  factors discussed above.
  Assuming a $\pi$-pulse (i.e. a pulse which leads to population inversion)
  with bandwidth covering all ground state
  vibrational levels, the maximum probability is given by 
  $|\langle\psi_g(t_D)|\psi_g(t_D)\rangle|^2$, restricting the sum in
  Eq.~(\ref{eq:psig_perturb}) to the bound levels. 
  The highest probability, 0.2,  is expected for a delay of 135 ps
  (cf. Fig.~\ref{fig:probability}), 
  when the wave packet $\Psi_{\mathrm{exc}}(R,t_P+135)$ is focussed after 
  half a vibrational period. Note the relatively short time window in which the probability 
  is maximum. A secondary maximum is visible 
  after one vibrational period (250 to 300 ps) when the last levels are populated:  much less 
  population can then be transferred. Note also the revival at $t_D \sim 400$~ps, after one 
  and a half vibrational period, with a larger width.
  The advantage of focussing is further  manifested by comparing to the Franck-Condon overlaps
  of wave packets created by the 
  pump pulses ${\cal P}_+ $ and ${\cal P}_0 $. The maximum probability  is now less than 10\%, 
  instead of 20\% for the focussed wave packet, and it is  spread during a larger time window.  
\item Next, we  report time-dependent 
  calculations  for various  dump pulses where the field is treated non-perturbatively
  and we start from a focussed wave packet. Intensity and spectral width are varied
  (see below Sec.~\ref{subsec:max})
  in order to check the validity limits of the above mentioned simple 
  model which relies on the impulsive approximation and weak fields. 
  Then a chirp of the dump pulse is introduced to obtain a robust scheme for 
  population inversion (cf. Sec.~\ref{subsec:chirp}). The dependence on 
  the time delay is studied in Sec.~\ref{ssec:timeup}.
  We furthermore present calculations  
  starting  from   vibrational wave packets created  with other  photoassociation pump pulses 
  such as ${\cal P_+} $, ${\cal P}_0 $  (Sec.~\ref{ssec:signchi}) and considering a  
  smaller detuning (Sec.~\ref{ssec:detun2}).
\end{itemize}

\section{Stabilization with a broadband pulse 
  populating many vibrational levels of the ground state}
\label{sec:many}

\subsection{Concept of a molecular $\pi$ pulse}
When the aim is to transfer as much population to the ground state as possible, the discussion 
of the pulse parameters is inspired by the arguments
on population inversion in molecular electronic states developed in 
Refs. \cite{cao98,vala01b}. The basic idea is that for pulses
which are short on the timescale of the vibrational dynamics,
the wave packet does not move during the pulse (impulsive approximation).
The molecular system can then be reduced to an effective two-level system, and it is possible 
to find a $\pi$-pulse which leads to total
population transfer. A $\pi$-pulse is defined as a  pulse (assumed here to be transform-limited
and with Gaussian envelope) whose
Rabi angle, 
%\begin{equation}
\begin{eqnarray}
  \label{eq:rabi}
  \vartheta &=& \int_{-\infty}^{+\infty} \mu \epsilon (t) dt =
  \int_{-\infty}^{+\infty} \mu \epsilon_0 \exp{\frac{-\sqrt{2 \ln 2}\;t^2}{\tau^D}} dt \\ \nonumber
  &=& \mu \sqrt{\frac{\pi}{\ln 2}} \tau^D\epsilon_0,  
\end{eqnarray}
%\end{equation}
is equal to an odd multiple of $\pi$. A pulse with  Rabi angle equal to an 
even multiple of $\pi$ induces zero population transfer.
Introducing a chirp enforces adiabatic following conditions for all
Rabi angles $>\pi$ and therefore
suppresses the oscillations in population transfer as a function of the Rabi
angle~\cite{cao98,vala01b,WrightPRL05}. 

\subsection{Inducing maximum population inversion}
\label{subsec:max}

The time-dependent Schr\"odinger equation describing the evolution of the wave 
packets $\Psi_{\mathrm{exc}}(R,t)$ and $\Psi_{\mathrm{g}}(R,t)$ in the excited and in the 
ground state, coupled by the radiative field of the dump pulse is solved numerically.
Our treatment of the field is non-perturbative (cf. Refs. \cite{luc2004a,luc2004b})
and for intensities corresponding to Rabi angles $\vartheta > \pi$
deviations from the  estimate of Eq.~(\ref{eq:psig_perturb}) are expected.
Deviations will also occur if the impulsive approximation ceases to be  valid, i.e. for longer
pulses. In this case, nuclear motion during the pulse cannot be neglected
and the wave packet is not simply projected to the electronic ground state.

For all calculations presented in this section, the parameters of the short dump pulses 
are recalled in Table~\ref{tab:dump_short}. The initial condition is given by
the focussed excited state wave packet at $t=t_P+135$~ps, normalized to one such that 
population corresponds directly to probability.  
The propagation is carried out in the time interval 
$[t_0=t_D -4 \sigma,t_\mathrm{final}=t_D +4 \sigma]$ with $\sigma= f_D\tau^D / 2\sqrt{2\ln2}$.
The  final population in the ground state $P_g(t_\mathrm{final})$ is given by
$P_g(t_\mathrm{final})= 
|\langle\Psi_{\mathrm{g}}(t_\mathrm{final})| \Psi_{\mathrm{g}}(t_\mathrm{final})\rangle|^2$ while the 
probability to form bound molecules corresponds to 
$P_g^\mathrm{bound}=\sum_i |\langle \varphi^g_i| \Psi_{\mathrm{g}}(t_\mathrm{final})\rangle |^2 $.
\begin{table*}[tb]
\vspace{0.5cm}
\begin{tabular}{|c|c|c|c|c|c|c|}
\hline\noalign{\smallskip}
$\omega_L^D$& $\hbar \delta \omega_D$&$\tau^D$& $f_D$&$ \chi_D$ & delay $t_D-t_P$ & Figs.\\
\noalign{\smallskip}\hline\noalign{\smallskip}
 11 843.5 cm$^{-1}$&2000-70  cm$^{-1}$ & 10-300 fs &1 & 0 & 135 ps &\ref{fig:rabi},\ref{fig:vibrabi} \\
\noalign{\smallskip}\hline\noalign{\smallskip}
 11 843.5 cm$^{-1}$& 200, 70  cm$^{-1}$   & 100, 300 fs & 2, 5, 10 & 120, 21.7, 5.52 ps$^{-2}$  & 135 ps &\ref{fig:rabichirp} to \ref{fig:overlapsign}\\
\noalign{\smallskip}\hline\noalign{\smallskip}
 11 845.0 cm$^{-1}$& 200, 70  cm$^{-1}$   & 100, 300 fs & 1 & 0 & 135 ps& \ref{fig:overlapdetun}\\
\noalign{\smallskip}\hline
\end{tabular}
\vspace{0.5cm} 
\caption{%
  Parameters for the short dump pulses   considered  in Sec. \ref{sec:many}: 
  central frequency $\omega_L^{at}$, temporal width $\tau^D$ (FWHME), stretching factor $f_D$, 
  linear chirp parameter $\chi_D$, delay between the maxima of dump and pump pulse $t_D-t_P $. 
  The corresponding 
  figures are reported in the last column.
}
\label{tab:dump_short}
\end{table*}

\begin{figure}[tb]
  \includegraphics[width=0.9\linewidth]{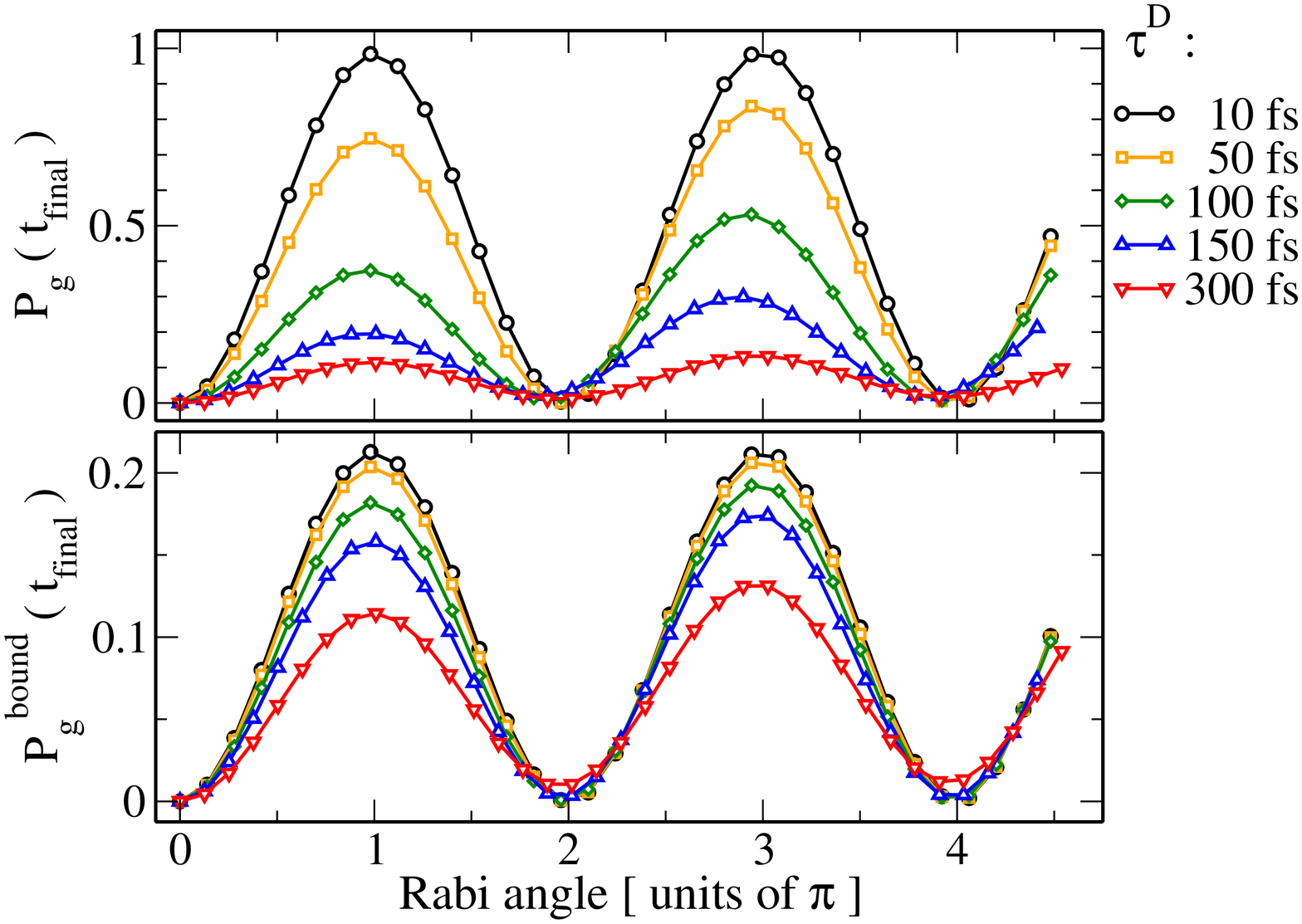}
  \caption{(Color online) %
    The final ground state population (top) and the final population  of bound
    levels in the ground state (bottom) after a transform-limited dump pulse 
    (see Table~\ref{tab:dump_short})
    as a function of the Rabi angle, i.e. intensity.
    For all pulse durations shown, we find pronounced Rabi oscillations, i.e.
    the approximation of an effective two-level systems holds. Very short pulses (10 fs)
    are able to induce complete population inversion, but the population which
    ends up in bound levels is limited to about 20\%; the dump pulse is 
    then dissociating 80\% of the photoassociated molecules. 
    Longer pulses (300 fs) populate only
    bound levels and no continuum states, since, within their bandwidth, 
    only bound levels are on resonance.
  }
  \label{fig:rabi}
\end{figure}

In Fig. \ref{fig:rabi}, we analyze the efficiency of the population transfer by varying the 
intensity of a transform-limited dump pulse for several durations $\tau^D$, in the range 
of 10-300 fs. The approximation of an effective two-level system is shown to hold for all pulse 
durations, since  oscillations as a function of the Rabi angle are clearly manifested in the 
final state population $P_g(t_\mathrm{final})$ and in the population of bound levels 
$P^{bound}_g(t_\mathrm{final})$.  The contrast of these oscillations is maximum for 
$\tau^D$=10 fs, where total population inversion is obtained with $P_g(t_\mathrm{final})$ 
oscillating between 0 and 1. For longer pulses ($\tau^D$=~300 fs), only  10\% of the 
population is transferred to the ground state. However, an analysis of the wave packet 
in the ground state shows that for the shorter pulses, the population is spread between 
continuum and bound levels. The probability to form bound molecules is at most 20\% 
($\tau^D$=10 fs), so that 80\% of the photoassociated molecules are dissociated by the 
dump pulse, giving pairs of hot atoms which usually leave the trap. With longer pulses, 
due to  their narrower frequency  bandwith,  the transfer to continuum levels is no longer 
possible:  for  $\tau^D$=300 fs the 11\% population transferred to the ground state is totally 
stabilized into bound levels.  The maximum population transfer and its efficiency are reported 
in Table~\ref{tab:pop}, demonstrating that in the intermediate range of pulse widths,
up to 100 fs, a stabilization of $\sim$ 20\% of the photoassociated molecules is to be expected.
\begin{table}[tb]
  \centering
  \begin{tabular}{|l|c|c|c|c|c|}
    \hline
    $\tau^D$ [ fs ] & 10     & 50     & 100    & 150    & 300 \\ \hline
    $P_{\mathrm g}$   & 0.984 & 0.746 & 0.374 & 0.195 & 0.114 \\ \hline
$P_{\mathrm g}^{\mathrm{bound}}$  & 0.212 & 0.208 & 0.180 & 0.158 & 0.114 \\ \hline
    $P_{\mathrm g}^\mathrm{bound}/P_{\mathrm g}$    & 0.216 & 0.273  & 0.487 & 0.811 & 0.999 \\
    \hline
  \end{tabular}
  \caption{ Ground state populations after a dump  $\pi$-pulse starting 
    from a unity normalized wave packet in the excited state;
    ratio of population in bound levels to overall ground state population
    after $\pi$-pulse (cf. Fig.~\ref{fig:rabi})}
  \label{tab:pop}
\end{table}

\begin{figure}[tb]
  \includegraphics[width=0.9\linewidth]{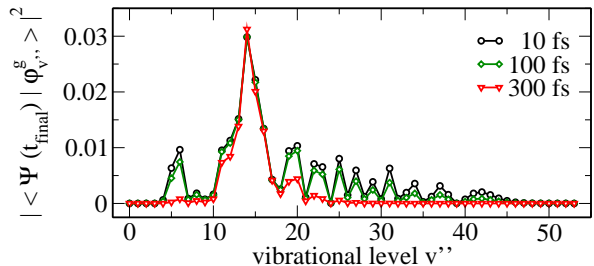}
  \caption{(Color online) The vibrational distribution of ground state wave packets
    after $\pi$-pulses (cf. Fig.~\ref{fig:rabi}):
    The distributions after a 10 fs and 100 fs
    pulse are almost identical  to the Franck-Condon overlap
    of the focussed excited state wave packet with ground state bound levels. 
    The narrower bandwidth of a 300 fs pulse allows for less levels to be populated 
    (cut-off around $v''=10$ and $v''$=25); 
    furthermore, no continuum states are excited (cf. Table~\ref{tab:pop}).}
  \label{fig:vibrabi}
\end{figure}

The possibility of stabilizing 20\% of the photoassociated molecules is very appealing. 
However, the vibrational distribution of the population transferred to the ground state 
should also be discussed, since stable molecules should be formed in low vibrational levels. 
The wave packet created in the ground state after a $\pi$ pulse is  analyzed through 
projection on the various $\varphi^g_{v''}(R)$ vibrational wave functions, cf.
Fig. \ref{fig:vibrabi}. For short enough  pulses ($\tau^D$ =10 fs, 100 fs) the 
distribution follows the variation of  the Franck-Condon overlap discussed above
(cf. Fig. \ref{fig:scheme}, right). The difference between the 10 fs and 100 fs
pulses affects mostly the population of continuum states 
(not shown in the figure). Due to the narrower bandwith, the longer pulse with $\tau^D$ =300 fs 
is only populating levels up to $v''$=25, and no continuum levels. 
In all cases the levels around $v''$=14 are preferentially populated.
The efficiency of the stabilization process   would of course be reduced  by considering 
Rabi angles different from $\pi$.  Therefore, for a given pulse duration,  the  scheme is 
very sensitive to the intensity of the dump pulse. 
This can be avoided by introducing {\em chirped} dump pulses~\cite{cao98,vala01b,WrightPRL05}.

\subsection{Chirping the dump pulse to suppress  oscillations as a function of the Rabi angle}
\label{subsec:chirp}
\begin{figure}[tb]
  \includegraphics[width=0.9\linewidth]{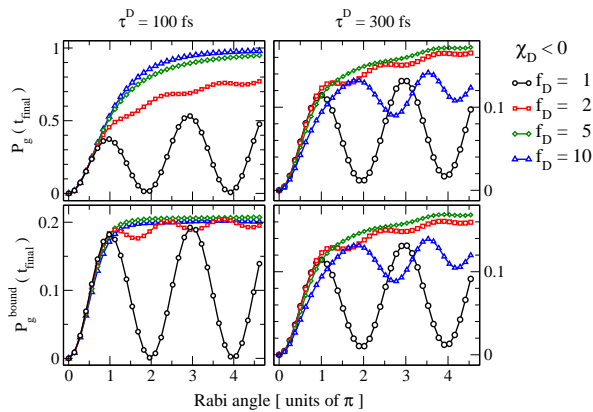}
  \caption{(Color online) %
    Overall ground state population (top) and population of bound ground state
    levels (bottom) after negatively chirped dump pulses 
    as a function of the Rabi angle
    for different values of the stretch factor $f_D$. 
    The duration of the transform-limited pulses is $\tau^D$= 100 fs (left)
    and $\tau^D$ = 300 fs (right). Results for
    an unchirped pulse are plotted for comparison (black circles).
  }
  \label{fig:rabichirp}
\end{figure}

In Fig. \ref{fig:rabichirp}, we report numerical results obtained after chirping the 
dump pulse, starting from transform-limited pulses of 100 fs and 300 fs, respectively.
The streching 
factor $f_D$ is varied from 1 to 10, and the linear chirp parameter $\chi_D$ is chosen
to be negative. Note that the Rabi angles in Fig. \ref{fig:rabichirp}
only characterize the pulse intensities, they are those of the corresponding unchirped pulses
(the Rabi angle becomes complex for chirped pulses). Fig. \ref{fig:rabichirp} clearly shows
that chirping suppresses the variation of the final population as a function of the Rabi angle.
For $\tau^D=$ 100 fs and $f_D \ge 5$
complete population inversion is reached at high intensities
due to non-perturbative effects such as power broadening (top left).
However, as for the unchirped pulses, the population of bound levels of the ground state 
remains limited to about 20\% (bottom left). The advantage of chirping
consists in this value of  20\% being obtained for a wide range of Rabi angles
($\pi$  to 4 $\pi$ in our calculations).
For $\tau^D$= 300 fs, nearly complete population inversion is 
achieved for stretching factors $f_D=2$, 5, and the probability of populating bound levels 
becomes close to 20\%, as for the shorter pulses. The chirp parameter should, however,
not be too large:  for a strongly chirped pulse ($f_D=10$) with corresponding
transform-limited FWHM $\tau^D$= 300 fs, 
the population transfer decreases, and oscillations reappear. This is due to the breakdown of the 
effective two-level system  approximation: With a   pulse  stretched to 3 ps,
it is no longer justified to neglect the relative  motion of the nuclei during the pulse.

\begin{figure}[tb]
  \includegraphics[width=0.9\linewidth]{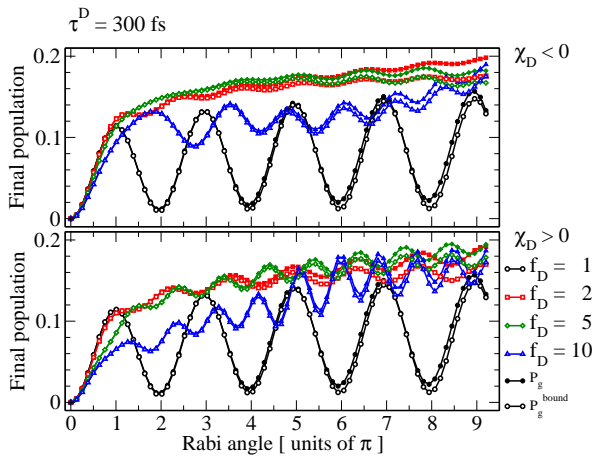}
  \caption{(Color online) Overall ground state population (filled symbols) and
    population of bound ground state levels (open symbols)
    after negatively (top) and positively (bottom) chirped dump pulses
    with transform-limited FWHM of 300 fs
    as a function of the Rabi angle
    for different values of the stretch factor $f_D$. The results for
    an unchirped pulse are plotted for comparison (black circles). 
  }
  \label{fig:rabichirp300fs}
\end{figure}

In Fig. \ref{fig:rabichirp300fs}, we
compare the final ground state population after chirped dump pulses with $\tau^D=  300$
for positive and negative chirp. The stretching factor is again varied from 1 to 10.
For pulses with corresponding transform-limited FWHM $\tau^D$= 100 fs, no difference between
negative and positive chirp was observed (data not shown). For  $\tau^D$=300 fs,
pulses with negative chirp
perform slightly better than those with positive chirp.
At high intensities ($\vartheta > 4\pi$), differences between open and
filled symbols are seen, i.e. even  fairly narrow-bandwidth
pulses start to excite continuum levels.

\begin{figure}[tb]
  \includegraphics[width=0.9\linewidth]{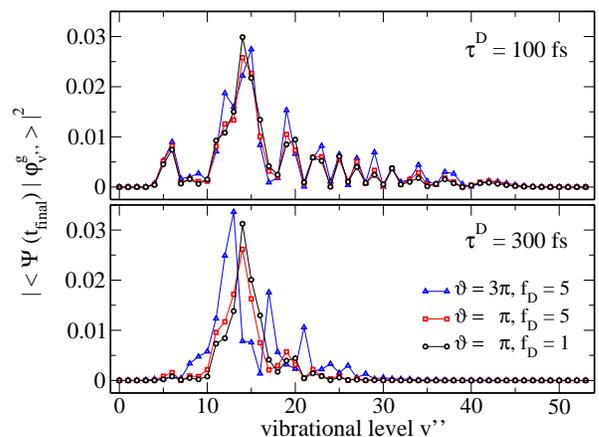}
  \caption{(Color online) The vibrational distribution of the final ground state
    wave packet after moderately chirped pulses ($f_D=5$, cf. Fig.~\ref{fig:rabichirp})
    with $\tau^D=100$ fs (top) and
    $\tau^D=300$ fs (bottom) for two different intensities corresponding
    to $\theta=\pi$ (red squares) and $\theta=3\pi$ (blue triangles).
    The vibrational distribution after an
    unchirped $\pi$-pulse is shown for comparison (black circles).
  }
  \label{fig:overlapchirp}
\end{figure}

The effect of the chirp of the dump pulse 
on the final vibrational distribution in the ground state is analyzed in
Fig. \ref{fig:overlapchirp}.
A chirped pulse with intensity corresponding to $\vartheta=\pi$
(red squares in Fig. \ref{fig:overlapchirp})
leads to very a similar vibrational distribution as an unchirped $\pi$-pulse
(black circles) for both pulse durations.
For higher intensities ($\vartheta=3\pi$, blue triangles), non-perturbative effects start to
play a role, and the vibrational distribution is changed: For example, 
the maximum is not anymore observed for the level with the largest
Franck-Condon factor ($v''=14$).
Furthermore for $\tau^D=300$ fs, the approximation of an
effective two-level system starts to break down due to the long pulse duration. 
While at higher intensity the dynamics becomes harder to interpret,
the overall probability to form bound molecules is
increased from 18\% ($\vartheta=\pi$) to
21\% ($\vartheta=3\pi$) for $\tau^D=100$~fs and from
11\% to 16\% for  $\tau^D=300$~fs.

\subsection{Dependence on the time spent on the excited state}
\label{ssec:timeup}
\begin{figure}[tb]
  \includegraphics[width=0.9\linewidth]{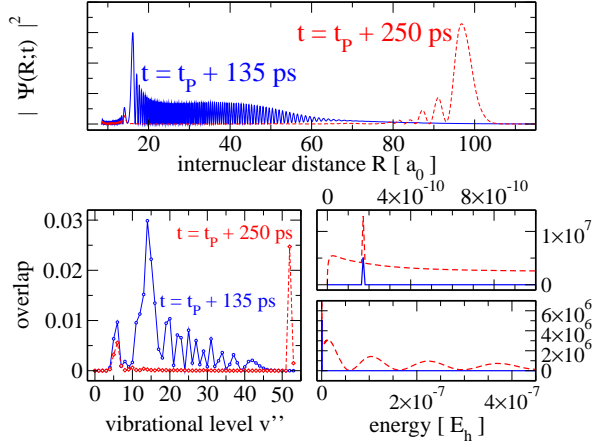}
  \caption{(Color online) top: The excited state wave packet for two different time delays
    between pump and dump pulses: $t=t_P+T_\mathrm{vib}/2$ (solid blue line)
    and  $t=t_P+T_\mathrm{vib}$ (dashed red line)
    ($t_P$ denotes the time where the pump pulse has its maximum).
    bottom: The Franck-Condon overlap of the
    excited state wave packets with the  ground state bound (left)  and the continuum
    (right) levels. 
  }
  \label{fig:excdyn}
\end{figure}

After excitation by the photoassociation pulse at long range, the wave packet
is accelerated toward shorter distances until it reaches the inner barrier
of the external potential well (solid blue line in Fig.~\ref{fig:excdyn}, top panel).
While a small part of the wave packet tunnels into
the inner potential well, most of it is reflected at the barrier
and moves back toward larger distances (dashed red line).
The Franck-Condon factors reflect this time-dependence of the
excited state wave packet (Fig.~\ref{fig:excdyn}, bottom panel, see also
Fig.~\ref{fig:schemenew}).
Short and sufficiently intense pulses ($\tau^D=100$ fs, $\vartheta=\pi$)
allow for inducing maximum population inversion in both cases,
and the final ground state vibrational distributions completely mimic the
Franck-Condon factors (data not shown).
However, only the wave packet at shorter distances has good Franck-Condon
overlap with bound ground state levels, and ground state molecules are formed
by the stabilization pulse. For the excited state wave packet at later times,
only the last two bound levels of the ground state are populated (except for
tunneling), and a considerable amount of population is transferred to the
continuum, i.e. two free atoms are created. Fig.~\ref{fig:excdyn} confirms our
initial assumption that stabilization to ground state molecules is optimal when
the excited state wave packet is close to the inner turning point of the
outer potential well.

\subsection{Dependence on the sign of the chirp of the photoassociation pulse}
\label{ssec:signchi}
\begin{figure}[tb]
  \includegraphics[width=0.9\linewidth]{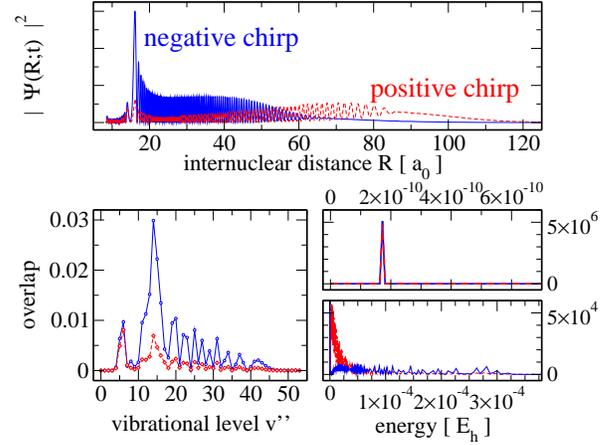}
  \caption{(Color online) top: The excited state wave packet at $t=t_P+135$~ps for
    positive and negative sign of the pump pulse chirp ($\cal{P}_-$ and $\cal{P}_+$).
    (bottom) The Franck-Condon overlap of the
    excited state wave packet with the  ground state bound  (left) and the continuum
    (right) levels. 
  }
  \label{fig:overlapsign}
\end{figure}

Excited state wave packets after pump pulses with positive ($\cal{P}_+$, red
dashed lines) and negative chirp ($\cal{P}_-$, blue solid lines)
are compared in Fig.~\ref{fig:overlapsign}
at the time of maximum focussing at the inner barrier.
The sign of the chirp determines how well the excited state wave packet can 
be focussed (cf. top panel of Fig.~\ref{fig:overlapsign}).
For positive chirp, the wave packet is less focussed, and the
maximum probability to form bound molecules is 6\% compared to 21\%
for negative chirp. Furthermore, almost half of these 6\% are due
 to the specific topology of the $0_g^-$-potential of Cs$_2$,
i.e. tunneling to the inner well.
Focussing as obtained with  $\cal{P}_-$ on the other hand 
represents a general scheme.
As in the previous sections the vibrational
distributions after the dump pulse (data not shown) completely mimic the Franck-Condon
overlap factors for pulses which induce maximum population transfer
(e.g. $\tau^D=100$ fs, $\vartheta=\pi$).
The population of bound levels is then 4.9\% for $\cal{P}_+$ to 18\% for $\cal{P}_-$.

\subsection{Dependence on the detuning of the photoassociation pulse %
  from the atomic line}
\label{ssec:detun2}
\begin{figure}[tb]
  \includegraphics[width=0.9\linewidth]{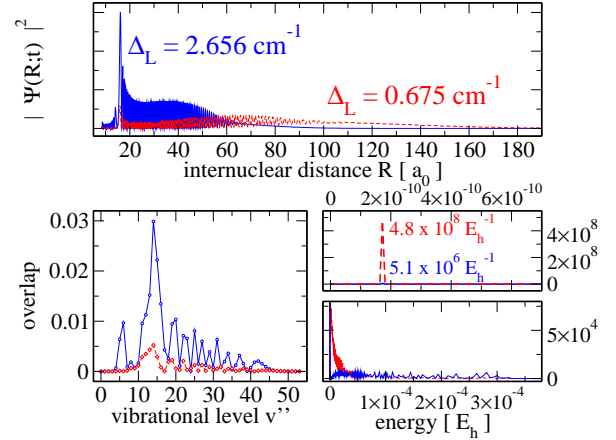}
  \caption{(Color online) top: The excited state wave packet at $t=t_P+T_\mathrm{vib}/2$ for
    two different detunings of the pump pulse.
    bottom: The Franck-Condon overlap of the
    excited state wave packet with the  ground state bound  (left) and the continuum
    (right) levels.
  }
  \label{fig:overlapdetun}
\end{figure}

Fig.~\ref{fig:overlapdetun} compares the excited state wave packet for
two different detunings of the pump pulse ($\cal{P}_-$ and ${\cal{P}}_-^{122}$,
respectively).
For smaller detuning (${\cal{P}}_-^{122}$),
higher vibrational levels of the excited state potential
get populated. Due to the larger spread of vibrational periods,
it is harder to fulfil the condition, Eq.~(\ref{eq:focus}), 
and the wave packet is less focussed
(cf. top panel Fig.~\ref{fig:overlapdetun}). Furthermore,
no tunneling to the inner well of the $0_g^-$ state occurs for 
smaller detuning. Hence, the peak
in the Franck-Condon overlap factors around $v''=6$ is missing.
Pulses which achieve maximum
population inversion to the ground state can still be found (e.g.
$\tau^D=100$ fs, $\vartheta=\pi$). However, 
the probability of populating bound ground state levels is at 3.4\%
considerably smaller.
The efficiency of the complete two-color scheme is of course given
by the efficiencies of pump {\em and} dump steps. One has to keep in mind
that since the probability of the photoassociation step is  higher
for ${\cal{P}}_-^{122}$ (see Table \ref{tab:pump}), the two effects compensate
and the final number of bound molecules is the same.

\section{How to create ground state molecules in  a single level}
\label{sec:single}

\begin{figure}[tb]
  \includegraphics[width=0.9\linewidth]{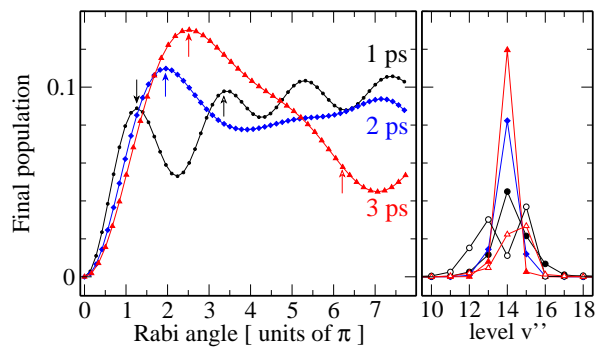}
  \caption{(Color online) left: Overall ground state population after transform-limited
    pulses with $\tau^D=1$ ps (black circles, 2 ps (blue diamonds), and 3 ps
    (red triangles) as a function of the Rabi angle.
    right: Vibrational distributions of the final ground state wave packet
    corresponding to the points indicated by arrows in the left panel.
  }
  \label{fig:single}
\end{figure}

\begin{figure}[bt]
  \includegraphics[width=0.9\linewidth]{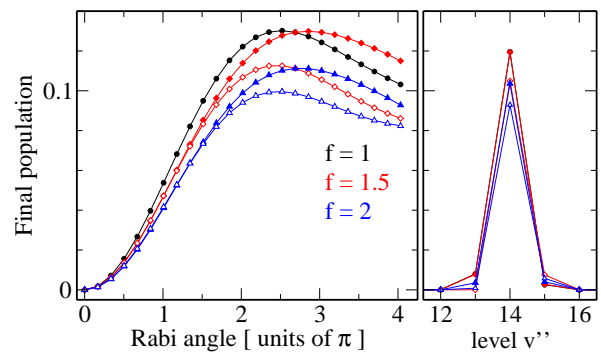}
  \caption{(Color online) left:  Overall ground state population after
    negatively (filled symbols) and positively (open symbols) chirped
    pulses with transform-limited FWHM of 3 ps as a function of the Rabi angle.
    The stretch factor is 1.5 (red diamonds) and 2 (blue triangles),
    respectively.
    The results for an unchirped pulse are shown for comparison (black circles).
    right: Vibrational distributions of the final ground state wave packet
    corresponding to the maxima of the final ground state population on the
    left-hand side.
    }
  \label{fig:single_chirp}
\end{figure}

In order to achieve selective population of the level $v''=14$,
we consider longer pulses, 
with  parameters  reported in Table~\ref{tab:dump_long}.
\begin{table*}[tb]
\vspace{0.5cm}
\begin{tabular}{|c|c|c|c|c|c|c|}
\hline\noalign{\smallskip}
$\omega_L$ & $\hbar \delta \omega$ & $\tau^D$ & $f_D$ &$\chi_D$ & delay $t_D-t_P$ & Fig. \\
\noalign{\smallskip}\hline\noalign{\smallskip}
 11 843.5 cm$^{-1}$ & 20-7 cm$^{-1}$ & 1, 2, 3 ps &  1      & 0                   & 135 ps & 13 \\
 11 843.5 cm$^{-1}$ & 7 cm$^{-1}$ & 3 ps       & 1.5, 2  & 1.84, 1.2 ps$^{-2}$  & 135 ps & 14 \\
\noalign{\smallskip}\hline
\end{tabular}
\vspace{0.5cm} 
\caption{%
  Parameters for the ``long'' dump pulses   considered  in Sec. \ref{sec:single}:
  central frequency $\omega_L^D$, temporal width (FWHME) $\tau^D$ 
  streching factor $f_D$, linear chirp parameter $\chi_D$, 
  delay between the maxima of dump and pump pulses. The corresponding figures are reported
 in the last column.}
\label{tab:dump_long}
\end{table*}
As explained in Sec.~\ref{ssec:param}, pulses with corresponding transform-limited
FWHM $\tau^D>2$ ps are expected to populate a single vibrational level. Fig. \ref{fig:single}
reports the final ground state population after dump pulses with  $\tau^D=1$ ps (black circles),
2 ps (blue diamonds) and 3 ps (red triangles) as a function of the Rabi angle, i.e.
pulse intensity on the left, and corresponding vibrational distributions on the right.
A pulse with $\tau^D=3$ ps and moderate intensity indeed excites exclusively
a single vibrational level (filled red triangles in Fig.~\ref{fig:single}, right).
The population of this level is $P_{v''=14}=0.12$ compared to the total
population of $P_g^\mathrm{bound}=0.13$.
A broadening of the vibrational distribution
is only observed at fairly high intensity ($\vartheta\sim 6\pi$,
empty red triangles). For $\tau^D= 1$ ps, broadening of the
vibrational distribution sets in already at lower intensities ($\vartheta\sim 3\pi$,
empty black circles).
For $\tau^D=1$ ps, oscillations of the final population with the Rabi angle 
still persist, albeit with much lower contrast
than for the broad-bandwidth pulses of Sec.~\ref{sec:many}. For
longer pulses, the impulsive approximation completely ceases to be valid, and
the behavior cannot anymore be understood (and employed for control)
in terms of an effective two-level system.
Chirping prolongs the
pulses even further, and its
effect on the final populations and vibrational distributions
is investigated in
Fig.~\ref{fig:single_chirp} which compares moderately chirped pulses
(stretching factors $f_D=1.5$, red diamonds and $f_D=2$, blue triangles)
to the unchirped case (black circles) for $\tau^D=3$ ps.
The final ground state population (blue triangles, $f_D=2$) is decreased after
a chirped pulse as compared to the unchirped case.
However, a slight negative chirp (filled red diamdonds, $f_D=1.5$)
does not significantly influence the population transfer.
In general, a range of intensities
corresponding to Rabi angles $2\pi\le\vartheta\le 3\pi$ exists
where 10\% or more of the
excited state wave packet are transferred to the ground state, almost
exclusively to the vibrational level $v''=14$ (cf. Fig.~\ref{fig:single_chirp}, right).
This result implies that some chirping (which might always occur in an experimental setup)
does not perturb stabilization into a single level, i.e. the scheme is expected to be
fairly robust.

%--------------------------------------------------------------------------------%
\section{Conclusions}
\label{sec:concl}
The formation of ultracold molecules is presently a very active field.
Previous work has demonstrated the advantage of using chirped laser pulses to
create ultracold molecules by photoassociation in an assembly of
cold atoms \cite{luc2004a,luc2004b}.
The molecules are created in an electronically excited state
and are short-lived. A stabilization step must therefore be implemented.
The present theoretical work  has
discussed a two-color pump-dump experiment where a second pulse, delayed from the
photoassociation (pump) pulse, transfers population to deeply bound levels
of the ground state, creating stable molecules.

Following Ref. \cite{luc2004a}, the pump pulse (with picosecond duration and
maximum at time $t_P$)  photoassociates Cs atoms  populating, via an
adiabatic transfer mechanism,  %$\sim$
about 15 vibrational levels in the external well of the %Cs$_{2}(
$0_g^-(6s\,\,^{2}S_{1/2}+6p\,\,^{2}P_{3/2}$) potential curve. A vibrational
wave packet is created which is bound by 2.1 cm$^{-1}$ and localized in the region of the
outer turning point. After the pulse, the wave packet  moves toward the inner
barrier of the potential curve (located at $R_\mathrm{in}$).
The chirp parameter of the pump
pulse can be chosen such that after half a vibrational period, at time $t_P$ + $T_\mathrm{vib}/2$,
this wave packet is focussed at the barrier \cite{luc2004a}.
Our analysis of the Franck-Condon overlap
integrals has shown that the overlap with deeply bound levels $v''$ of the ground triplet
state is then maximum. A vertical transition  toward the level $v''$=14,
bound by 113 cm$^{-1}$, is even possible, since the outer turning point
of the corresponding wave function coincides with $R_{in}$.
We therefore have discussed how  population can efficiently be transferred to the
ground triplet state by a short dump pulse which should be
in the femto- or picosecond domain and optimally delayed from the pump
pulse by $T_\mathrm{vib}/2$. % or other times.

The validity of this intuitive model has been checked with numerical calculations.
We have considered
two kinds of pulses: broadband pulses populating many vibrational levels of the ground
state, with the aim of maximizing the population transfer, and longer pulses for selective
population of  a single level,  $v''$=14.

The first set of calculations has been performed for a time delay of $T_\mathrm{vib}/2$.
We have shown that for short pulses and a Rabi angle equal to $\pi$,
total population
inversion can be obtained, i.e. the vibrational wave packet in the excited state is
completely transferred to the ground state.
Analysis of the wave packet shows that a maximum
of 20\% of this population goes to bound levels. The remaining 80\% correspond
to continuum states, i.e. the dump pulse also creates pairs of hot atoms. The
vibrational distribution reflects the variation of  the Franck Condon overlap.
Oscillations in the population transfer as a function of the Rabi angle
(i.e. laser intensity for a fixed pulse duration) can be
suppressed by chirping the dump pulse. Thus,  total population
inversion is obtained for a wide range of intensities
(corresponding to Rabi angles of $\pi$  to $4\pi$ in our calculations)
with an effective 20\% population transfer to bound levels.
Employing a chirped dump pulse
  therefore yields a robust scheme in which the probability to create
stable ground state molecules is 20\%, independent of the exact value of laser intensity.

By varying the time-delay between the two pulses,  a strong variation in the
population transfer which follows  the variation of the Franck-Condon factors
has been observed.
Only  at shorter distance, the wave packet in the excited state  has good Franck-Condon
overlap with bound ground state levels, and ground state molecules are formed
by the dump  pulse. The highest efficiency is
achieved in a narrow time window ($T_\mathrm{vib}/2\pm 2$ps) during which the
wave packet is maximally focussed at the barrier.
For different time-delays, 
only the last two bound levels of the ground state are populated, and a considerable
amount of population is transferred to the
continuum, i.e. two free atoms are created.
Note that without a second pulse, stabilization
would occur via spontaneous emission on a nanosecond timescale. The probability
for creating ground state molecules is then given by a time average. Since the wave packet
spends more time at large distances than close to the barrier, times with unfavorable
Franck-Condon overlaps dominate this average. 
\textit{It is therefore essential to employ a second pulse to fully take advantage of the
  time-dependent probability to create bound ground state molecules after
  photoassociation with picosecond pulses.} 

The calculations have confirmed our
intuition  that stabilization to ground state molecules is optimal when
the excited state wave packet is close to the inner turning point of the
outer potential well. The best localization of the wave packet at the inner barrier
is brought about by negatively chirping the pump pulse. There is yet another advantage 
when using a negative chirp of the pump pulse: 
The photoassociation probability is increased  
by a factor of 8 as compared to a transform-limited pulse.
Furthermore, the probability of creating bound molecules by the dump pulse is
enhanced by roughly a factor of 3 for a negatively chirped pump pulse as compared
to  positively chirped or  transform-limited ones. 
The number of stable molecules
formed in a pump-dump sequence can be therefore be
increased by 24 when appropriately choosing the pulse parameters.

We have also considered a pump pulse with smaller detuning, for which the photoassociation
probability is larger. However, due to a decrease of the Franck-Condon overlap, even for
the maximally focussed wave packet, the stabilization
step is much less efficient. Therefore, the total number of stable molecules is roughly
the same as in the previous case.
 
Finally, we have employed longer dump pulses with a duration of about 3 ps
to selectively populate
the $v''$=14 level in the ground state. Up to 12\% of the photoassociated molecules can
then be stabilized into a \textit{single} level, which is fairly efficient.
Note that these molecules are created coherently by the pump-dump sequence.

The full discussion of the definition of the optimal pulse should include the possibility 
of a repetition scheme, 
considering further pump and dump pulses. In treating a repetition scheme,
we also have to take into account how the pump and dump  pulses modify the continuum wave function
describing two colliding atoms. This is beyond the scope of the present study.

In conclusion, a two-color pump-dump scheme is 
rather promising for the creation of stable cold molecules.
Due to the choice of the Cs$_2$ 0$_g^-$ potential
for the photoassociation step , the present work has
analyzed formation of molecules in the $v''$=14 level of the ground triplet state,
bound by 113 cm$^{-1}$. Many other schemes, e.g.
\cite{SagePRL05,KochPRA04}, are possible and should be considered to
bring population into the  $v''$=0 level of the ground state
which will open the way to the formation of stable molecular
condensates. In this respect, we would like to point out the generality of the
two-color pump-dump scheme
which only requires efficient photoassociation and stabilization mechanisms
involving either long range wells in the excited potentials~\cite{FrancoiseReview}
or resonant coupling~\cite{dion01,dulieu03,kerman04},
but is not restricted to a particular system.

\begin{acknowledgments}
  We would like to acknowledge fruitful discussions with Ronnie Kosloff and to
  thank Anne Crubellier for computing independently overlap integrals with a
  Numerov method in order to check our numerical results.
  This work has been supported by
  the Deutsche Forschungsgemeinschaft (C.P.K.) and 
  the European Union in the frame of the 
  Cold Molecule Research Training Network under contract HPRN-CT-2002-00290.
  The Fritz Haber Center is supported
  by the Minerva Gesellschaft f\"{u}r die Forschung GmbH M\"{u}nchen, Germany.
  Laboratoire Aim\'e Cotton belongs to F\'ed\'eration Lumi\`ere Mati\`ere
  (CNRS, Universit\'e Paris XI).
\end{acknowledgments}

\appendix

\section{A note on the representation of continuum states
  with a mapped Fourier grid}
\label{sec:box}

The Fourier grid employs a finite number of grid points and periodic
boundary conditions to represent a quantum wave function.
For $V(R) \equiv 0$, this corresponds to the particle in a box problem.
The scattering continuum is hence represented by a finite number of
discrete box states. The highest momentum which is correctly represented
on the grid is given in terms of the (constant) spacing $\Delta R$ between two
grid points, $p_\mathrm{max}=\pi/\Delta R$.

In order to model long-range interactions which require large grids but not
small grid steps, a mapping procedure is introduced.
The grid steps in the asymptotic region become very large, correspondingly
high momenta are cut off. This alters the representation of continuum states
on the grid. When diagonalizing the Hamiltonian,
two kinds of continuum eigenfunctions are obtained: true
box levels and wave functions with an unphysical shape. The latter will be called
``cutoff-states'' in the following.

The box levels can be easily identified by the scaling of their eigenenergies,
$E_n \sim n^2$ (except for the first few states just above the dissociation limit
where the energies are perturbed by the threshold).
The corresponding wave functions are standing waves of the box.

The eigenenergies of the cutoff-states approximately scale
with some high power ($\sim 20$) of the box index, the highest eigenenergy
being determined by the largest allowed grid step which is a parameter of the mapping.
The corresponding eigenfunctions show very fast oscillations and don't
obey the boundary conditions at $R_\mathrm{min}$ and $R_\mathrm{max}$. It can nevertheless
be shown that these cutoff-states constitute an \textit{effective} representation
of the true, physical continuum for energies higher than the energy of the last
true box level. To this end, the Franck-Condon overlaps of an excited state wave function
with the eigenfunctions of all ground state levels (i.e. bound, box and cutoff-levels)
are calculated and compared (i) for different mapping parameters and (ii)
to Franck-Condon overlaps obtained by Numerov integration \cite{crubellier2005}.

In the first set of calculations, the largest allowed grid step, $\Delta R_\mathrm{max}$,
is varied. The Hamiltonian is diagonalized for each $\Delta R_\mathrm{max}$ and an
increasing number of grid points, $N_R$. For each $\Delta R_\mathrm{max}$,
the number of cutoff-states turns out to be
independent of $N_R$, i.e. increasing $N_R$ leads to
a larger number of true box states. In particular, for the smallest
$\Delta R_\mathrm{max}$ and largest $N_R$ used
($\Delta R_\mathrm{max}=0.33$~a$_0$, $N_R=8191$),
a continuum representation is obtained which contains almost no cutoff-states
(168 vs. 7969 true box states).
The excited state wave function is then interpolated
to these grids, and the Franck-Condon overlaps are computed.
Note, that the Franck-Condon overlaps for continuum states need to include the
correct energy renormalization~\cite{luc2004b}.
The Franck-Condon overlaps turn out to be roughly independent of the grid parameters.
We are particularly interested in the sum over the Franck-Condon overlaps which
determines the probability of dissociating the excited state molecules
into hot, free atoms with a short dump pulse (cf. Sec.~\ref{sec:many}).
This sum is roughly constant for all grids considered,
varying between 0.21 for $\Delta R_\mathrm{max}=33$~a$_0$ and
0.17 for $\Delta R_\mathrm{max}=0.33$~a$_0$. The variation is caused by
the interpolation of the excited state wave function to the new grids
which is not exact, the approximation getting worse for decreasing
$\Delta R_\mathrm{max}$.

In a second set of calculations, Franck-Condon overlaps with all
ground state levels were calculated for 
selected excited state vibrational levels with both
the mapped Fourier grid and the Numerov method.
Oscillations like those in Fig.~\ref{fig:excdyn} are observed. The resolution
of these oscillations in the mapped Fourier grid method
depends on the number of grid points, $N_R$.
Good agreement between the two methods is obtained for $N_R=1023$, the value
used throughout Secs.~\ref{sec:many} and \ref{sec:single}.

%--------------------------------------------------------------------------------%

\end{document}